\documentclass[psfrag,aps,prb,twocolumn,showpacs,floatfix,%
superscriptaddress,amssymb,amsmath]{revtex4}

\usepackage{graphicx}
\usepackage{dcolumn}% Align table columns on decimal point
\usepackage{bm}% bold math
\usepackage{amsmath}
\newcommand{\dlangle}{\langle\langle}
\newcommand{\drangle}{\rangle\rangle}
\usepackage{color}
\usepackage{psfrag}
\begin{document}

\title{Phonon-assisted tunneling regimes in diatomic molecules}

\author{E. Vernek}
\affiliation{Department of Physics and Astronomy, and Nanoscale
and Quantum Phenomena Institute, \\Ohio University, Athens, Ohio
45701-2979} \affiliation{Departamento de F\'{\i}sica,
Pontif\'{\i}cia Universidade Cat\'olica, Rio de Janeiro-RJ,
Brazil}

\author{E. V. Anda}
\affiliation{Departamento de F\'{\i}sica, Pontif\'{\i}cia
Universidade Cat\'olica, Rio de Janeiro-RJ, Brazil}

\author{S. E. Ulloa}
\affiliation{Department of Physics and Astronomy, and Nanoscale
and Quantum Phenomena Institute, \\Ohio University, Athens, Ohio
45701-2979}

\author{N. Sandler}
\affiliation{Department of Physics and Astronomy, and Nanoscale
and Quantum Phenomena Institute, \\Ohio University, Athens, Ohio
45701-2979}

\date{\today}

\begin{abstract}
Electronic transport in diatomic molecules (two-level systems) connected 
to metallic contacts is analyzed in the presence of competing electron-electron
and electron-phonon interactions. We show that phonon emission and absorption
processes are strongly modified when a Coulomb energy $U$ is included, as the phonons open channels that can result in destructive or constructive interference effects. Resonance conditions
for these processes produce dramatic effects both in the density of states at the molecular sites, as well as in the conductance through the system. We find in particular an
enhanced {\it Rabi-assisted tunneling} due to phonons, as the resonance conditions are met, which is made more evident for increasing temperatures.  These effects are controllable by voltage gating of the molecular sites, and should be accessible in current experiments.
\end{abstract}

\pacs{73.23.--b, 73.63.Kv, 71.38.--k, 73.63.--b}
\keywords{Coulomb blockade, molecular
electronics, phonon scattering, polarons}
%Use showkeys class option if keyword display desired

\maketitle

\section{introduction}

Electronic transport properties of nanometer scale structures such as natural and artificial molecules (e.g. quantum dot structures) attached to metallic reservoirs have received a great deal of attention
recently.  Three main reasons drive the active research in the area: i) these systems have a fundamental role in the development of the field of ``molecular electronics" where molecules serve as electronic components;\cite{james} ii) they are relatively ``simple" structures in which to address different conceptual issues regarding the nature of charge and spin transport; \cite{spins} and iii) they represent the prototypical gateable structures at the nanoscale. \cite{rmp-gates}

One of the most defining characteristics of these molecular systems is the strong spatial confinement that results in enhanced electron-electron interaction (EEI), and profoundly modifies their transport properties. Well studied examples of the  effect of EEI in confined systems are Coulomb blockade and Kondo effect in quantum dots. \cite{liang,jeong,glazman1,glazman2,martinek,aguado2}
Molecular systems may also exhibit strong electron-phonon interactions (EPI), which play a significant role in their physical properties. Several works have addressed the effects of EPI in quantum-dot molecules for different geometrical arrangements, in a perturbative regime. \cite{stauber,bissiri,li}
It is known for example that phonons cause a broadening of Coulomb blockade peaks and the appearance of satellite peaks in the nonlinear transport regime. Phonons also provide relaxation mechanisms (through inelastic scattering processes) that affect the electronic conductance, specially in molecular systems, {\cite{ortner,verzelen,gartner}} as well as polaronic shifts of the electronic levels. \cite{Aleiner}

While the effect of these different interactions on electronic transport
properties has been studied extensively in regimes where one of them is
dominant, not much is known when both interactions become comparable. Recently, two works have addressed this issue. Al-Hassanieh {\it et al}. have studied the effect of electron-phonon
interactions in the transport properties of a single-level quantum dot connected to two metalic contacts in the Kondo regime. \cite{busser}
They found an unexpected cancelation of the conductance resulting from interference between the pure electronic and phonon-assisted tunneling channels.

Our group reported on the effect of competing interactions
on the conductance of a diatomic molecule coupled to metallic reservoirs in the Coulomb blockade regime and for temperatures above the Kondo scale. \cite{Vernek}  That work focused on a regime
where the energy difference between the two local electronic levels is much larger than the broadening of the levels due to the coupling to the leads. In that case, indirect coupling between the local levels is negligible and interference of the different transport channels from one lead to another is not important. Utilizing the equations of motion for the various Green's functions involved, we found a splitting in the electronic density of states (DOS) of the molecule under suitable resonance conditions.  This effect was identified as {\it Rabi-assisted tunneling}, because of the formal similarity with the well-known Rabi splitting phenomena in atomic systems, \cite{Boca} although in this case it is due to the interplay of EEI and EPI.\@ The physical origin of the Rabi-assisted tunneling can be traced back to the appearance of new channels for electron tunneling between the metallic reservoirs. These channels, due to EEI, provide new resonance conditions for phonon emission and absorption. An immediate consequence of this result is the possibility of important changes in the conductance as function of the gate potential that controls electron occupation in the molecule, and which would be evident in experiments. This surprising result warrants a closer examination of the approximations used in the solution of the model and a thorough analysis of the various resonance and interference conditions stemming from the EPI in the Hamiltonian.

In the present paper, we address these issues and include a complete analysis of the effects of the indirect coupling between the two electronic levels induced by the metallic leads.  We now explore the regime where the local electronic level spacing tends to zero, which makes the phase effects much more important in the coherent transport processes. We also study in detail all resonant regimes and conducting channels provided by the EPI.\@  To solve the equations of motion, we use a Hubbard-like decoupling procedure for all the Green's functions involved in the calculation. \cite{hubbard1} This approach has proven reliable in the description of properties of  interacting systems at temperatures above their characteristic Kondo temperature.  These calculations allow us to confirm the existence of the Rabi-splitting in the DOS and the appearance of resonances and anti-resonances not obtained previously. We show that there are structures in the DOS depending upon resonance and anti-resonance conditions whose activation is level-occupation--dependent, reflecting the many-body nature of the problem.

We further explore how the DOS structures and resonance conditions affect the conductance through the diatomic molecule structure.  We find that different Coulomb blockade peaks in the absence of EPI, appear split whenever phonons are present and this effect is enhanced at higher temperatures.

\section{Model and theoretical Method}

\begin{figure}
%\vskip-.6cm
\includegraphics{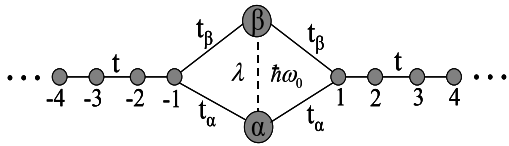} 
 \caption{\label{Fig1} Schematic representation of the model system.
Only the electron-phonon interaction connects the local sites
($\epsilon_\alpha^0 < \epsilon_\beta^0$) via phonons of frequency
$\omega_0$ and with coupling constant $\lambda$.}
\end{figure}

Our model consists of a diatomic molecule with a single level per site  (or equivalently a quantum dot pair) with local energies $\epsilon_{\alpha}^0$ and $\epsilon_{\beta}^0$ (we assume the non-degenerate case with $\epsilon_{\beta}^0 > \epsilon_{\alpha}^0$), as schematically shown in Fig.\ \ref{Fig1}.

Each single-level site (or quantum dot) is connected independently to two external metallic leads. We include local (on-site) electron-electron interactions and an electron-phonon inter-site interaction. As described the model can represent various experimental geometries; in particular, this model was studied in the limit of non-interacting electrons, to explore interference effects. \cite{eto} The total Hamiltonian is then written as
\begin{eqnarray}
H_T &=& H_{mol}+H_{leads}+H_{mol-leads} \, , 
\end{eqnarray}  
where
\begin{eqnarray} 
H_{mol} &=& H_{el}+H_{ph}+H_{el-ph} \, , 
\end{eqnarray}
with
\begin{eqnarray}
H_{el}&=&\sum_{\sigma,i=\alpha,\beta}\left[\epsilon_i c^{\dagger}
_{i\sigma}c_{i\sigma} + \frac{U}{2} n_{i\sigma}
n_{i\bar{\sigma}}\right] , \\
H_{ph}&=&\left(b^{\dagger}b+\frac{1}{2}\right)\hbar\omega_0 \, ,
\end{eqnarray}
and the electron-phonon interaction is given by
\begin{equation}
H_{el-ph}=\lambda\sum_{\sigma}\left(b^{\dagger}
c^{\dagger}_{\alpha\sigma}c_{\beta\sigma}+h.c\right) \, ,
\end{equation}
where $b^{\dagger}$ ($b$) creates (annihilates) a phonon with energy $\hbar\omega_0$, $\epsilon_i=\epsilon^0_i-eV^i_g$, and $i = \alpha, \beta$. We have assumed that $U_\alpha = U_\beta = U$, and included independent gate voltages $V_g^i$ that control the electron occupation of the molecular sites by shifting the localized energies with respect to the Fermi energy of the left and right electrodes. Also,
\begin{eqnarray}
H_{mol-leads}=\sum_{i=\alpha,\beta \atop \gamma=R,L } t_{i\gamma}
c^{\dagger}_{i\sigma} c_{\gamma k\sigma}+h.c.
\end{eqnarray}
and 
\begin{eqnarray}
H_{leads}=\sum_{ k,\sigma\atop \gamma=R,L} \epsilon_{\gamma k}
c^{\dagger}_{\gamma k\sigma} c_{\gamma k\sigma}.
\end{eqnarray}
$H_{mol-leads}$ connects the diatomic molecule to the leads; for simplicity, we consider $t_{i\gamma}=t^{\prime}$, for all $\gamma=R,L $ and $i=\alpha,\beta$.  Away from the Kondo regime, the main effect of the leads is to broaden the energy levels of the dots through the tunnel couplings $t^{\prime}$.  To describe the transport of electrons through the molecule, we calculate the local retarded Green's
functions  defined in the usual way as, \cite{Mahan,zubarev}
\begin{eqnarray}
\langle\langle A;B\rangle\rangle &\equiv& \int i\theta (t) \langle
[A(t),B(0)]_{\mp}\rangle e^{i\epsilon t}dt \, ,
\end{eqnarray}
where $A$ and $B$ are operators and $\langle \cdots\rangle$ denotes the thermodynamic average (for $T\neq 0$) or the ground state expectation value (for $T=0$). In the frequency domain, the equation of motion for Green's functions is written as
\begin{eqnarray}\label{eom}
\omega \langle\langle A;B\rangle\rangle =
\langle[A,B]_{\pm}\rangle+\langle\langle[A,H]_-;B\rangle\rangle,
\end{eqnarray}  
where the subindex in the brackets denotes commutation ($-$) and anti-commutation ($+$) relations. The expression for the local electronic Green's function is:
\begin{eqnarray}
G^\sigma_{ii}(\omega)&=&\dlangle c_{i\sigma};c^\dagger_{i\sigma}\drangle,
 \ \ \ \ \ \ \ \ \hbox{$i=\alpha,\beta$} \, .
\end{eqnarray}

The assumption of identical leads allows the introduction of symmetric and antisymmetric  fermionic operators $c_{Sk\sigma}$, $c_{Ak\sigma}$ 
\begin{subequations}
\begin{eqnarray}
c_{Sk\sigma}&=&\frac{1}{\sqrt{2}}(c_{Rk\sigma}+c_{Lk\sigma})\\
c_{Ak\sigma}&=&\frac{1}{\sqrt{2}}(c_{Rk\sigma}-c_{Lk\sigma}).
\end{eqnarray}
\end{subequations}
After this transformation,  $H_{mol}$ and $H_{mol-leads}$ read as
\begin{eqnarray}
\tilde H_{leads}=\sum_{k,\sigma \atop {\gamma=S,A}}  \epsilon_{k}
c^{\dagger}_{\gamma k\sigma} c_{\gamma k\sigma}.
\end{eqnarray}
and
\begin{eqnarray}
\tilde H_{mol-leads}=\sum_{k,\sigma \atop i=\alpha,\beta} \tilde t^\prime
c^{\dagger}_{i\sigma} c_{S k\sigma}+h.c.\,,
\end{eqnarray}
where we have assumed $\epsilon_{Rk}=\epsilon_{Lk}=\epsilon_k$,  and $\tilde t^\prime=\sqrt{2}t^\prime$. In the symmetric-antisymmetric basis, the coupling between the molecule and the leads occurs only through the symmetric channel.  The other terms in the Hamiltonian are not affected by this transformation.  We obtain the relevant Green's functions by using the equation of motion technique. As an illustrative example, let us describe the steps taken in the calculation of the Green's function for level $\alpha$ :
\begin{eqnarray}
(\omega - \epsilon_\alpha) \dlangle c_{\alpha\sigma};c^\dagger_{\alpha\sigma}\drangle&=& 1+ U\dlangle
n_{\alpha\sigma}c_{\alpha\sigma};c^\dagger_{\alpha\sigma}
\drangle \nonumber \\ && +\lambda\dlangle b^\dagger
c_{\beta\sigma};c^\dagger_{\alpha\sigma}\drangle\nonumber \\
&&+ \sum_{k} {\tilde t^\prime} \dlangle c_{Sk\sigma};c^\dagger_{\alpha\sigma}\drangle \, .
\end{eqnarray}
To obtain a closed set of equations, we apply a Hubbard-like decoupling procedure. In particular, for
Green's functions that involve operators for both levels $\alpha$ and $\beta$ (``mixed Green's functions''), we close the equations by considering the following approximations: 
\begin{subequations}
\begin{eqnarray}
 \langle\langle b^\dagger
c^\dagger_{\alpha\bar\sigma}c_{\beta\bar\sigma}c_{\alpha\sigma}
 ;c^\dagger_{\alpha\sigma}\rangle\rangle&\approx&\langle b^\dagger
c^\dagger_{\alpha\bar\sigma}c_{\beta\bar\sigma}\rangle\langle\langle
c_{\alpha\sigma}
 ;c^\dagger_{\alpha\sigma}\rangle\rangle\\
\langle\langle b
c^\dagger_{\beta\bar\sigma}c_{\alpha\bar\sigma}c_{\alpha\sigma}
 ;c^\dagger_{\alpha\sigma}\rangle\rangle&\approx&\langle b
c^\dagger_{\beta\bar\sigma}c_{\alpha\bar\sigma}\rangle\langle\langle
c_{\alpha\sigma}
 ;c^\dagger_{\alpha\sigma}\rangle\rangle\\
 \langle\langle b^\dagger b c_{\alpha\sigma}
 ;c^\dagger_{\alpha\sigma}\rangle\rangle&\approx&\langle b^\dagger
b\rangle\langle\langle
 c_{\alpha\sigma} ;c^\dagger_{\alpha\sigma}\rangle\rangle \, .
\end{eqnarray}
\end{subequations}
The first two equations are the standard Hubbard decoupling procedure. At this level of the approximation, correlations between electron operators with opposite spins are neglected. The third equation decouples the phonon mode from the electron operators, as suggested by Zubarev.\cite{zubarev} Finally, by keeping up to second order terms in $\lambda$ in the expression of the self-energies, we obtain:
\begin{eqnarray}
G_{ii}(\epsilon) = \left\{%\frac{1}{
\frac{\left(\epsilon-\epsilon_i\right)\left(\epsilon-\epsilon_i-U\right)
}{\left[\epsilon-\epsilon_{i}-U\left(1-\langle
n_{i\bar\sigma}\rangle\right)\right]}-
\Sigma_{i\sigma}(\epsilon) \right\}^{-1}%}
\label{greenf},
\end{eqnarray}
with  $\Sigma_{i\sigma}(\epsilon)=\Sigma^{(el)}_{i\sigma}(\epsilon)
+\Sigma^{(ph)}_{i\sigma}(\epsilon)$. Here $\Sigma^{(el)}_{i\sigma}(\epsilon)$ is the electron self-energy in level $i$ due to the leads,  and $\Sigma^{(ph)}_{i\sigma}(\epsilon)$ is the phonon-related self-energy.  They have the following expressions:
\begin{widetext}
\begin{eqnarray}
\Sigma^{(el)}_{i\sigma}(\epsilon)&=&  \tilde t^{\prime 2}\tilde
g(\epsilon)\left\{ 1+\frac{\left[
\epsilon-\epsilon_j-U\left(1-\langle
n_{j\bar\sigma}\rangle\right)\right]\tilde t^{\prime 2}\tilde
g(\epsilon)}{
\left(\omega-\epsilon_j\right)\left(\epsilon-\epsilon_j-U\right)
-\left[\epsilon-\epsilon_j-U\left(1-\langle
n_{j\bar\sigma}\rangle\right)\right]\tilde t^{\prime 2}\tilde
g(\epsilon)} \right\}\label{Sel},\\
\Sigma_{\alpha\sigma}^{(ph)}(\epsilon)&=&\frac{\lambda^2\left[\langle b^\dagger
b\rangle+
 \langle n_{\beta\sigma}\rangle\right]
\left[\epsilon-\epsilon_\beta+\hbar\omega_0-U(1-\langle
n_{\beta\bar\sigma}\rangle) \right]}{
\left(\epsilon-\epsilon_\beta+\hbar\omega_0\right)\left(\epsilon-\epsilon_\beta-
U+
\hbar\omega_0\right)-\left[\epsilon-\epsilon_\beta+\hbar\omega_0-U(1-\langle
n_{\beta\bar\sigma}\rangle) \right]\tilde t^{\prime
2}\tilde
g(\epsilon+\hbar{\omega_0)}}\label{S1ph},\\
\Sigma^{(ph)}_{\beta\sigma}(\epsilon)&=& \lambda^2\frac{\left[\langle b^\dagger
b\rangle+1-\langle
n_{\alpha\sigma}\rangle\right]
\left[\epsilon-\epsilon_\alpha-\hbar\omega_0-U(1-\langle
n_{\alpha\bar\sigma}\rangle)\right] }{
\left(\epsilon-\epsilon_\alpha-\hbar\omega_0\right)\left(\epsilon-
\epsilon_\alpha-
U-\hbar\omega_0\right)-\left[\epsilon-\epsilon_\alpha-\hbar\omega_0-U(1-\langle
n_{\alpha\bar\sigma}\rangle)\right]\tilde t^{\prime2}\tilde
g(\epsilon-\hbar\omega_0)  }.\label{S2ph}
\end{eqnarray}
\end{widetext}
In all these expressions $\tilde g(\epsilon)$ refers to the Green's function of the leads at energy $\epsilon$ (both leads have the same Green's function in equilibrium). In Eq.\ (\ref{Sel}), the expression for the self-energy $\Sigma^{(el)}_{i\sigma}(\epsilon)$ of an electron in level $i$, involves energies and occupation numbers of electrons in the {\em other} level $j$. These occupation numbers  $\langle n_{j\sigma} \rangle$ have to be calculated self-consistently.  Since we are interested in the non-magnetic regime, the occupation numbers are
taken to be spin-independent ($\langle n_{j\bar\sigma}\rangle=\langle
n_{j\sigma}\rangle$) and they will be denoted just by $\langle n_j\rangle$.  Notice that the electron self-energy includes the {\em indirect coupling} between the two levels due to the presence of the leads through the second term inside the brackets in the denominator of Eq.\ (\ref{Sel}). The relevance of this term reduces as the
inter-level separation in the dots increases. As a consequence, the net effect of the electron self-energy is to produce a finite width in the DOS peaks, to shift their positions by an amount ${\tt Re}\Sigma^{(el)}$, and to give rise to transport interference effects due to the topology of the system.

Equations (\ref{S1ph}) and (\ref{S2ph}) are the phonon self-energies calculated up to second order in the electron-phonon interaction strength $\lambda$. They involve phonon occupation numbers $\langle b^\dagger b\rangle$, which for simplicity can be calculated from the free-boson Hamiltonian. Although this approximation is not essential, it is convenient to simplify the numerical calculation, and it does not introduce any artifact in the results, as it is a frequency independent $c$-number.

\begin{figure}
\resizebox{3.0in}{!}{\includegraphics{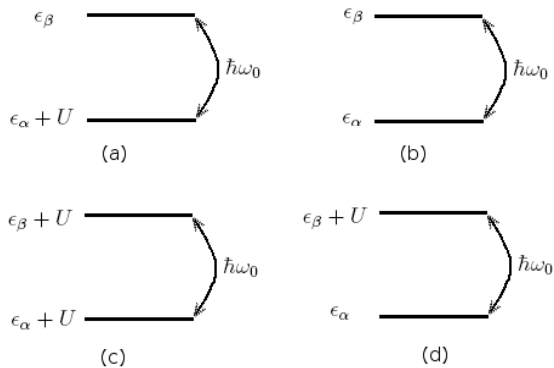}}
\caption{\label{ressonances} Schematic
representation of the various resonances and the active levels. The
figures correspond to the resonance conditions (a) \ref{resonance1}, (b) and
(c) \ref {resonance2} and (d) \ref{resonance3}.}
\end{figure}

From Eq.\ (\ref{greenf}), we obtain the total DOS by using
$\rho_\sigma(\epsilon)=(-1/\pi)[{\tt Im}G^\sigma_{\alpha\alpha}(\epsilon)+{\tt
Im}G^\sigma_{\beta\beta}(\epsilon)]$. In the limit where the electron-phonon
coupling is negligible ($\lambda \to 0$), the DOS presents four Coulomb blockade
peaks coming from the original atomic levels located at
$\epsilon_\alpha$, $\epsilon_\alpha+U$, $\epsilon_\beta$ and $\epsilon_\beta+U$.
As the electron-phonon interaction is turned on, these peaks show Rabi
splittings, at characteristic energies, corresponding to phonon-absorption
and/or phonon-emission processes at resonance conditions. These resonance
conditions occur when the phonon energy $\hbar\omega_0$ matches the energy
difference between the two energy levels.  These resonance conditions occur
when 
\begin{subequations}
\begin{eqnarray}
% \begin{alignedat}
\hbar\omega_0+U=\Delta\epsilon\label{resonance1}\\
\hbar\omega_0=\Delta\epsilon \label{resonance2}\\
\hbar\omega_0-U=\Delta\epsilon \, ,
\label{resonance3}
% \end{alignedat}
\end{eqnarray}
\end{subequations}
where $\Delta \epsilon = \epsilon_{\beta} - \epsilon_{\alpha}$.
Figure \ref{ressonances} shows a schematic representation  of the effective
levels active in each resonance.  Notice that the resonance conditions do not explicitly depend on the electron occupation number (unlike the simpler approximation results in [\onlinecite{Vernek}]).
However, when the system is found in one of these resonances the splitting 
$\delta_\alpha\propto {\tt Re}\Sigma^{(ph)}_{\alpha}$ for level $\alpha$
(and $\delta_\beta\propto{\tt Re}\Sigma^{(ph)}_{\beta}$ for level $\beta$),
depends on the gate voltage through the occupation numbers.

Using the expressions above for the Green's functions, we proceed to calculate the effect of electron-phonon interactions on the DOS and the conductance of the system. Transport properties are calculated in linear response in the fully
interacting regime ($U \neq 0$). The conductance  has
the following expression in the Keldysh formalism: \cite{Mahan} 
\begin{eqnarray}
G=4\pi^2t^4\rho_L(\omega)\rho_R(\omega)|G_{\bar1 1}(\omega)|^2\big|_{
\omega=\epsilon_F }. \label{condeq}
\end{eqnarray}
where
\begin{eqnarray}\label{cond}
G_{\bar1 1}(\omega)=[\tilde g(\omega)t^\prime]^2\left [
G_{\alpha\alpha}+G_{\beta\beta}+G_{\alpha\beta}
+G_{\beta\alpha}\right] 
\end{eqnarray}
is the Green's function that describes the dynamics of an electron with energy
$\omega$ from the site $-1$ to $1$ (representing the leads, see
Fig.~\ref{Fig1}) and $\tilde g(\omega)$ is the Green's function of the
leads, calculated as disconnected from the molecule. As the
four terms of Eq.\ (\ref{cond}) are complex quantities with different phases, their
contributions to the current can interfere constructively or destructively as
we discuss below. The off-diagonal Green's functions ($G_{\alpha \beta}$ and $G_{\beta \alpha}$) in the conductance are also calculated by equation of motion techniques at the same level of approximation described above. Finally, notice that we have neglected inelastic processes in the present calculation because we assume that the system is essentially at thermodynamical equilibrium, at the small bias voltages of interest.

\section{Numerical results}

Equations (\ref{greenf})-(\ref{S2ph}) are solved numerically and used to calculate the DOS and conductance in (\ref{condeq}), as
described above. For convenience, the Fermi energy is
set to zero ($\epsilon_F=0$). In what follows,  all the physical quantities are
given in units of the inter-level separation $\Delta \epsilon$. The levels
$\epsilon^0_\alpha$ and $\epsilon^0_ \beta$ are set at $1.0$ and $2.0$,
respectively (above $\epsilon_F$). The effect of an increase in the gate voltage
$V_{g}$ is lowering these levels (initially towards the Fermi energy).  In the next few figures, we assume equal gating of both sites, so that $V_g^\alpha =V_g^\beta = V_g$.

\begin{figure}
\begin{center}
%\begin{tabular}{cc}
 \resizebox{3.82in}{!}{\includegraphics{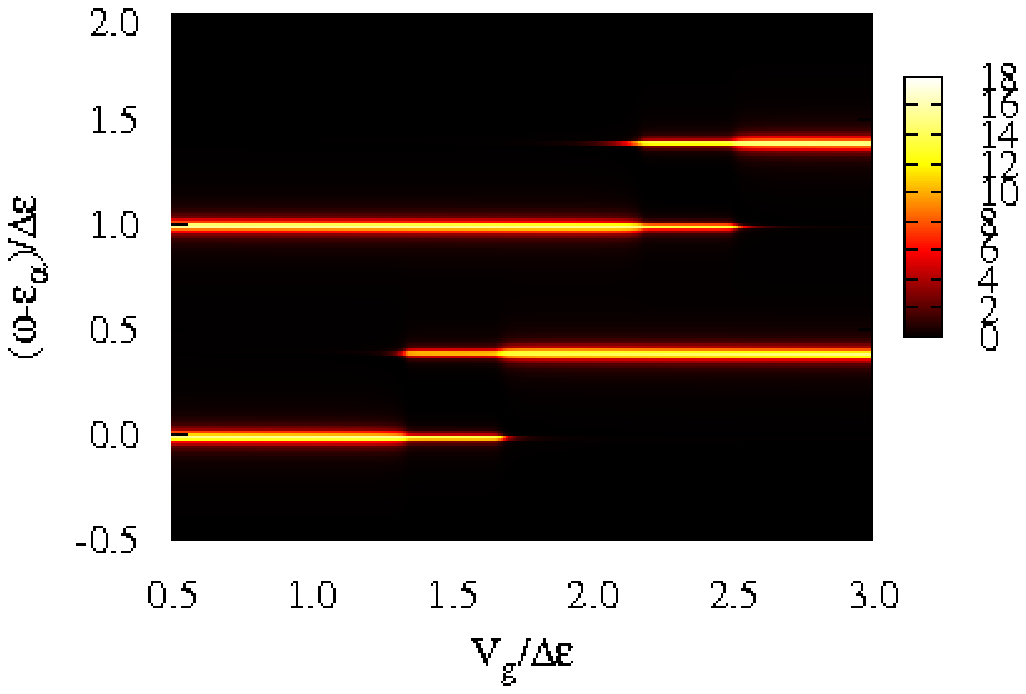}} \\ 
\hspace*{3ex} \resizebox{3.7in}{!}{\includegraphics{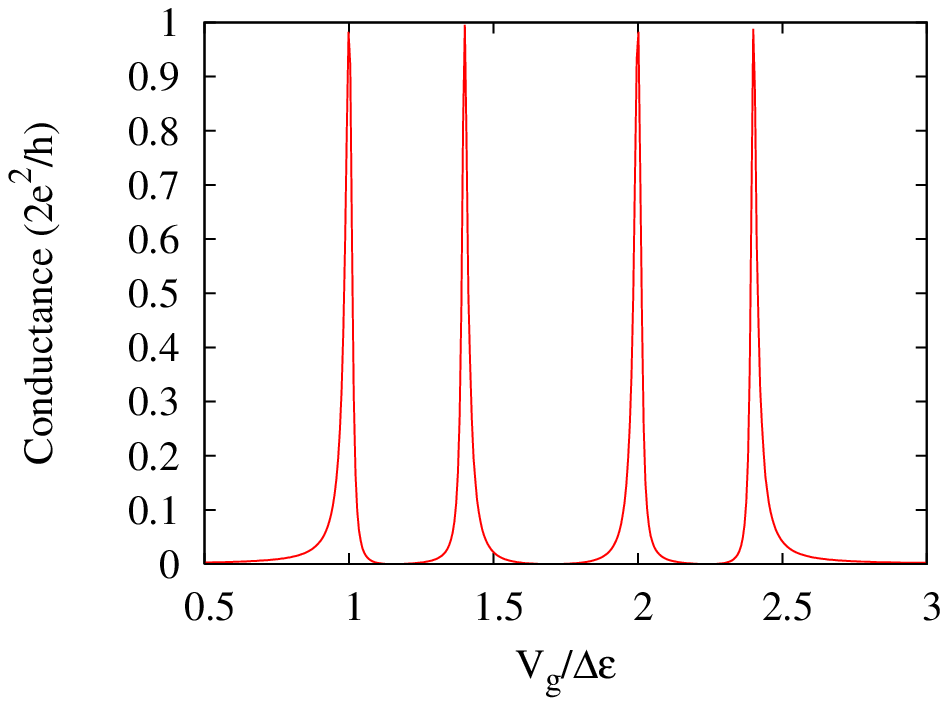}}
%\end{tabular}
\end{center}
\caption{(Color online) Density of states map as function of gate voltage and
energy (top panel) and conductance as function of gate voltage (bottom) for a
molecule with electron-electron interaction, but no electron-phonon interaction.
In units of $\Delta\epsilon = \epsilon_\beta - \epsilon_\alpha $, the parameters
are $\lambda=0$, $U=0.4$, $\hbar\omega_0=0.6$, $\tilde t^\prime = \sqrt{2}/5$, $\epsilon_\alpha^0 = 1$, and $T=0.0025$. Notice energy scale in top panel is shifted by $\epsilon_\alpha$.} 
\label{condl0} 
\end{figure}

Let us consider first the results for the interacting system in the absence of 
electron-phonon ineraction ($U \neq 0; \lambda = 0$). Figure (\ref{condl0}) shows
the results for the DOS (top) and the conductance (bottom) for the particular
values $U = 0.4$ and $T = 0.0025$ (energies in units of $\Delta \epsilon= \epsilon_\beta - \epsilon_\alpha$). As expected, the conductance exhibits peaks located at gate voltages near $\epsilon_\alpha$, $\epsilon_\alpha+U$, $\epsilon_\beta$ and $\epsilon_\beta+U$; these are the well-known Coulomb blockade peaks that appear whenever each level is in resonance with the Fermi energy.  Notice slight shifts from the anticipated energy values, caused by the non-vanishing real part of the electron self-energy discussed above.

\begin{figure}
\begin{center}
%\begin{tabular}{cc}
\resizebox{3.2in}{!}{\includegraphics{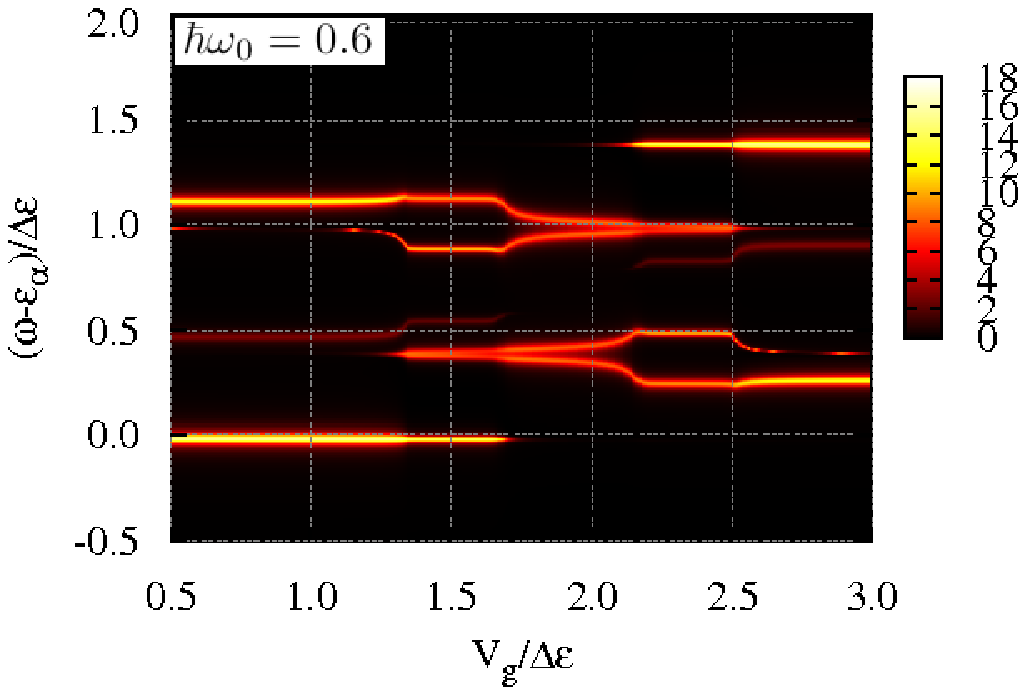}} \\
% \vskip-0.75cm
\resizebox{3.2in}{!}{\includegraphics{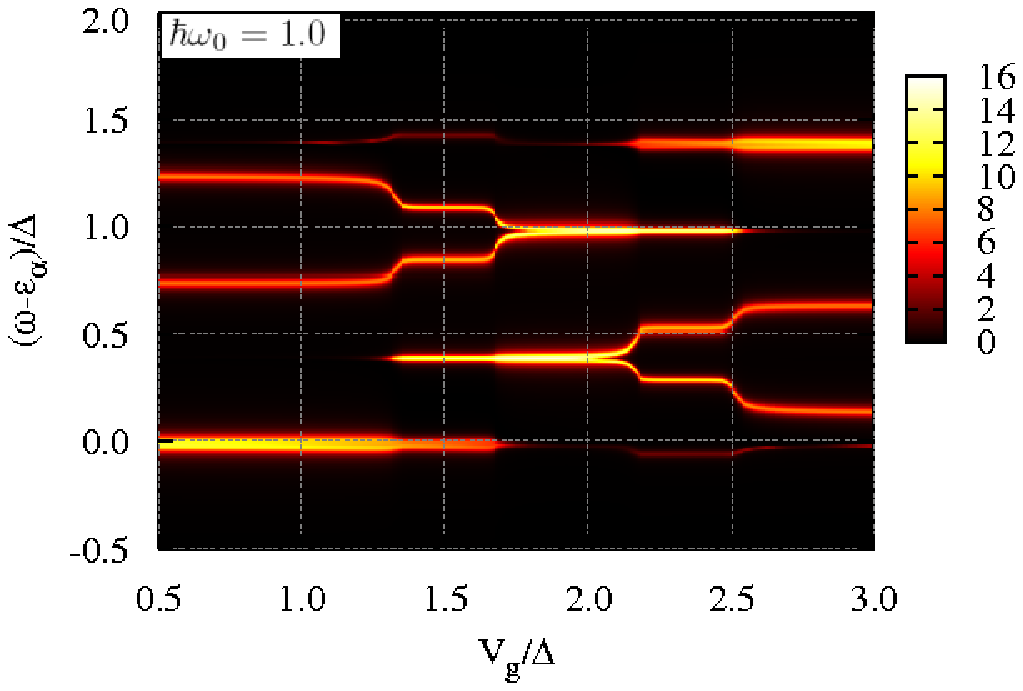}}\\
\resizebox{3.2in}{!}{\includegraphics{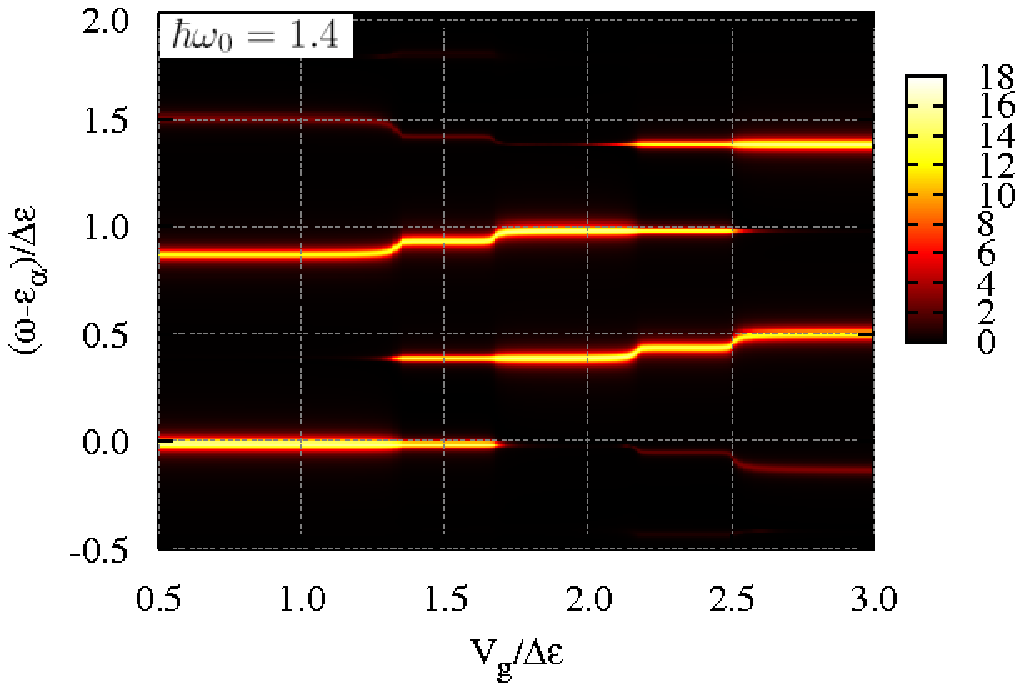}}
%\end{tabular}
\end{center}
\caption{\label{dos1} (Color online) Map of the DOS as function of gate voltage and
energy in the EEI and EPI competing regime. From top (in units of $\Delta\epsilon$), with $\hbar\omega_0=0.6$, $\hbar\omega_0=1$ and $\hbar\omega_0=1.4$.  In all panels, $U=0.4$, $\tilde t^\prime = \sqrt{2}/5$, $\lambda =0.2$, $\epsilon_\alpha^0 = 1$, and $T=0.0025$. The energy axes are shifted by $\epsilon_\alpha$, so that level $\alpha$ appears always at zero.}
\end{figure}

\begin{figure}
\begin{center}
%\begin{tabular}{cc}
  \resizebox{3.4in}{!}{\includegraphics{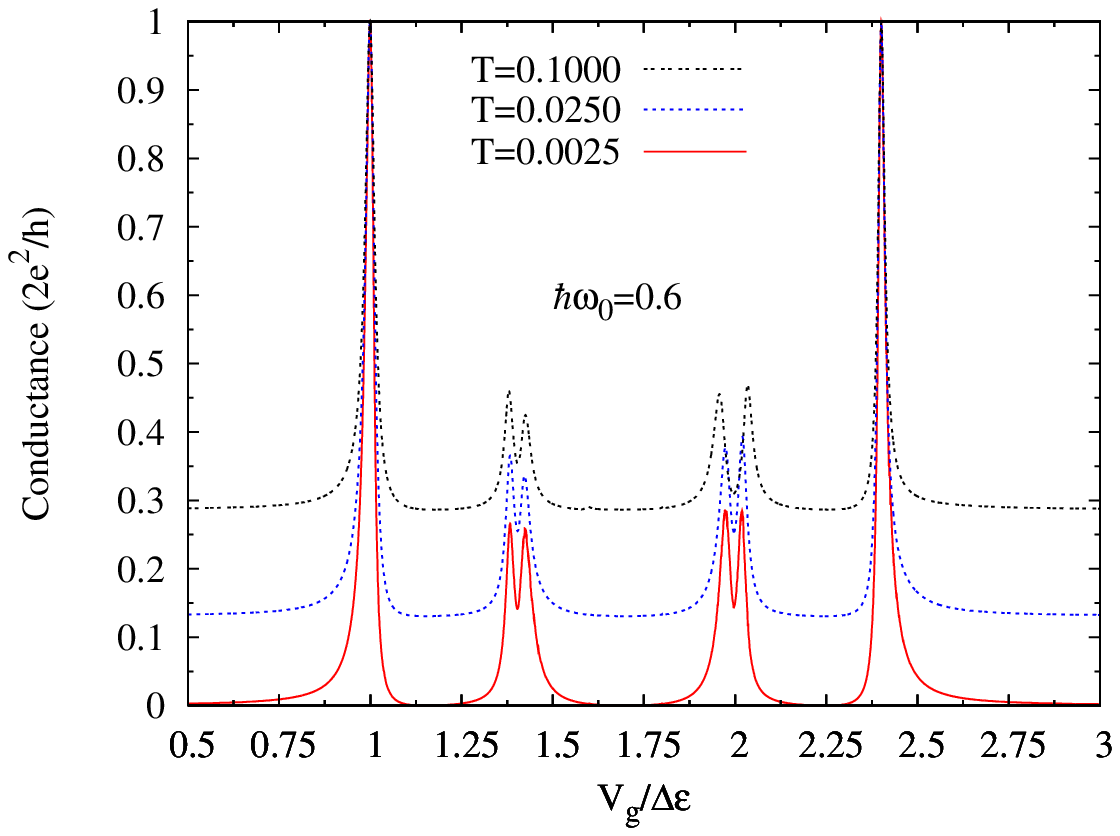}}\\
  \resizebox{3.4in}{!}{\includegraphics{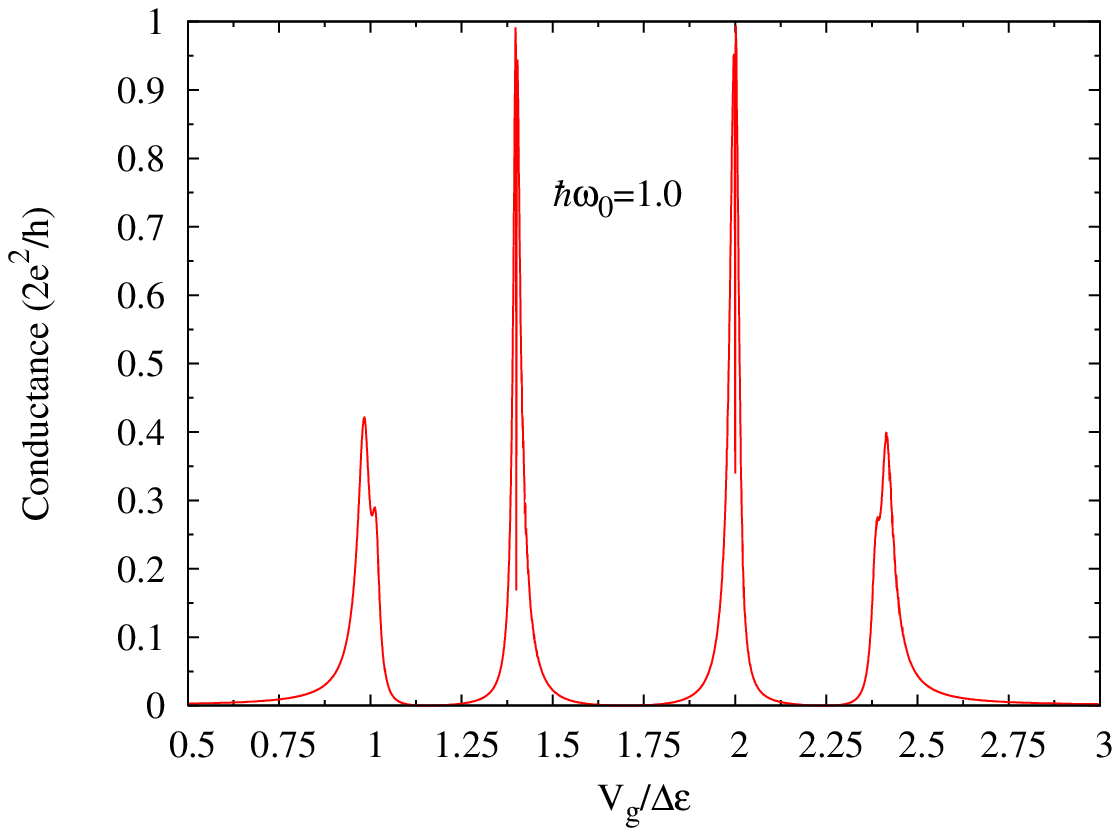}}\\
  \resizebox{3.4in}{!}{\includegraphics{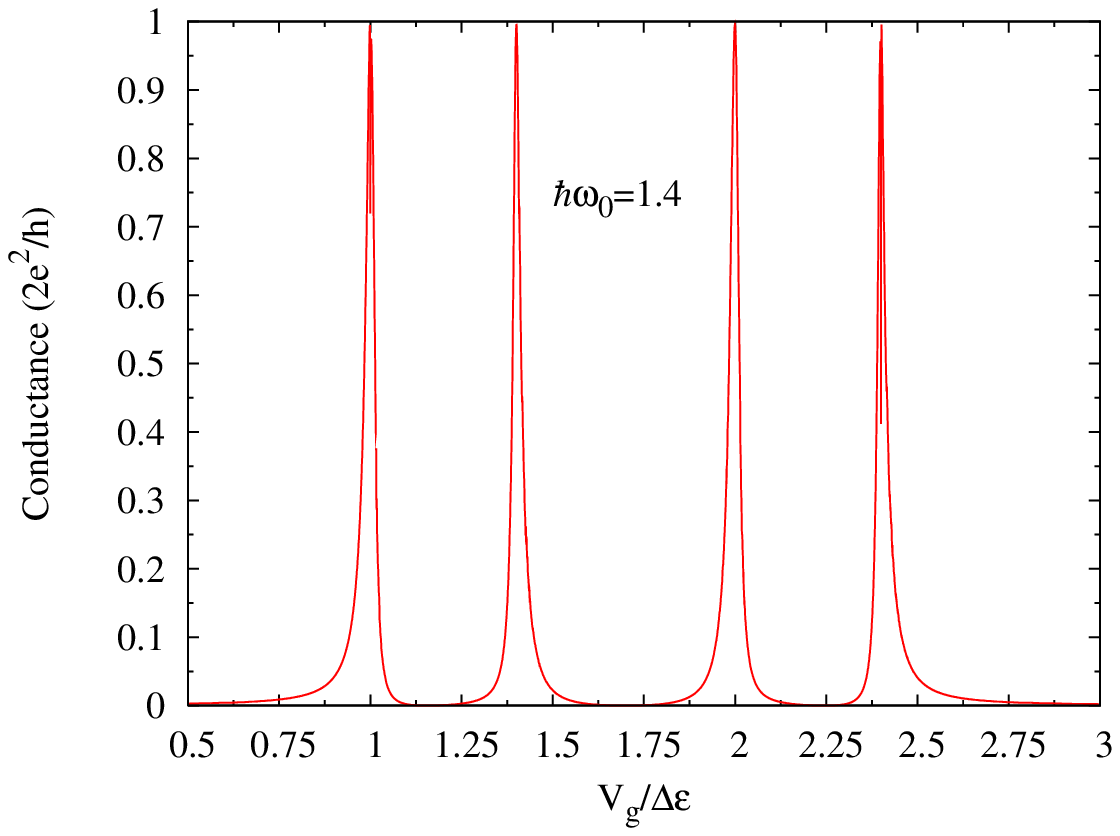}}
%\end{tabular}
\end{center}
\caption{\label{cond1} (Color online) Conductance as function of gate voltage for system as in Fig.\ \ref{dos1}. From top to bottom, $\hbar\omega_0=0.6$,
$\hbar\omega_0=1$ and $\hbar\omega_0=1.4$; parameters as above.  
Top panel also shows $T$ dependence of conductance for $\hbar \omega_0 = 0.6$;
notice splittings in middle two peaks {\em increase} with increasing
temperature. In the top panel the curves are offset for clarity.}
\end{figure}

Figure \ref{dos1} shows the DOS as a function of gate voltage $V_g$ and
energy $\omega$ in the presence of electron-phonon interaction. The different
plots refer to different values of the phonon energy $\hbar \omega_{0}$. In all
cases, the electron-electron interaction is set at $U = 0.4$, the
electron-phonon coupling strength $\lambda = 0.2$, and $T = 0.0025$. 
The values for the phonon energies, $\hbar\omega_0=0.6, 1.0, 1.4$, have been
chosen to fit the resonance conditions given by Eqs.\ (\ref{resonance1}),
(\ref{resonance2}) and (\ref{resonance3}), respectively.
From the figure, it can be observed that in the first resonance condition for
the smaller $\hbar\omega_0$ ($=0.6$, top panel) the peaks corresponding to the
levels $\epsilon_\alpha+U$ and $\epsilon_\beta$ are split, mixed by the phonon
channels. The  splitting for the peaks corresponding to the level
$\epsilon_\alpha +U$ is $\approx 2\delta_\alpha$ ($\approx 0$ 
at a gate voltage $V_g\approx1.4$), while the splitting for the level $\epsilon_\beta$ is $\approx 2\delta_\beta$ ($\approx 0.05$ at  $V_g \approx 2.0$, for example). 
Notice that an analysis of Eqs.\ (\ref{Sel})-(\ref{S2ph}), reveals that when level
$\epsilon_\alpha$ coincides with $\epsilon_F$ the splitting $\approx
2\delta_\alpha$ becomes small. This can be understood in terms of the occupation
number for the level $\beta$ which in these conditions remains close to zero
(this is seen by analyzing the value of the ${\tt Re} \Sigma_\alpha$, which  approaches zero). A similar analysis shows that when the level $\epsilon_\beta$
crosses $\epsilon_F$ the splitting $2\delta_\beta$ is also small, since ${\tt Re} \Sigma_{\beta}\approx 0$, a consequence of the levels $\epsilon_\alpha$ and $\epsilon_\alpha+U$ being far below $\epsilon_F$ (almost full levels). 
The splitting of the levels in the DOS is clearly, and most importantly, manifested in the conductance as a function of the gate voltage $V_{g}$, as shown in Fig.\ \ref{cond1}.

In the case of the second resonance condition, $\hbar\omega_0=1.0$ (middle panel
in Fig.\ \ref{dos1}),  there are two pairs of levels involved: $\epsilon_\alpha$, $\epsilon_\beta$ and $\epsilon_\alpha+U$, $\epsilon_\beta+U$. The effect of the resonance conditions on the conductance results to be much less important than in the previous case, as the splitting in the DOS at the Fermi level is much smaller here.  Notice, however, that in this case all four conductance peaks exhibit a certain degree of splitting (see middle panel in Fig.\ \ref{cond1}), indicating that the interference effects involve two pairs of levels (as in Fig.\ \ref{ressonances}b and \ref{ressonances}c).

For the last resonance condition, $\hbar\omega_0=1.4$, the levels
involved are $\epsilon_\alpha$ and $\epsilon_\beta+U$. These levels are quite far
apart, separated by $\Delta \epsilon + U$, which results in the splittings in the DOS  and the conductance to be barely observable, except for the sharp dip/splitting in the first and last Coulomb blockade peaks (lower panel, Fig.\ \ref{cond1}). 
We notice that the Rabi splitings are generally less important for the last two resonance conditions. The reason for this behavior is that larger $\hbar\omega_0$ means that the original levels participating in the phonon processes are far apart, decreasing the spliting energies $\delta_\alpha$ and $\delta_\beta$. This is clearly reflected in the respective conductance curves, where the peak splittings are very small.

We have also analyzed the dependence of these results on temperature. The top panel in Fig.\ \ref{cond1} shows the conductance as a function of gate voltage
for three different temperatures, $T=0.0025$, 0.025, and 0.1. A relatively weak
dependence on temperature is observed, consistent with the fact that the
resonance conditions are not explicitly dependent of the electron occupations. 
However, the self-consistency of the calculation induces an indirect occupation dependence, resulting in splittings that become more pronounced for increasing temperature ({\em i.e}., larger phonon numbers).

\begin{figure}
\begin{center}
\begin{tabular}{cc}
  \resizebox{3.9in}{!}{\includegraphics{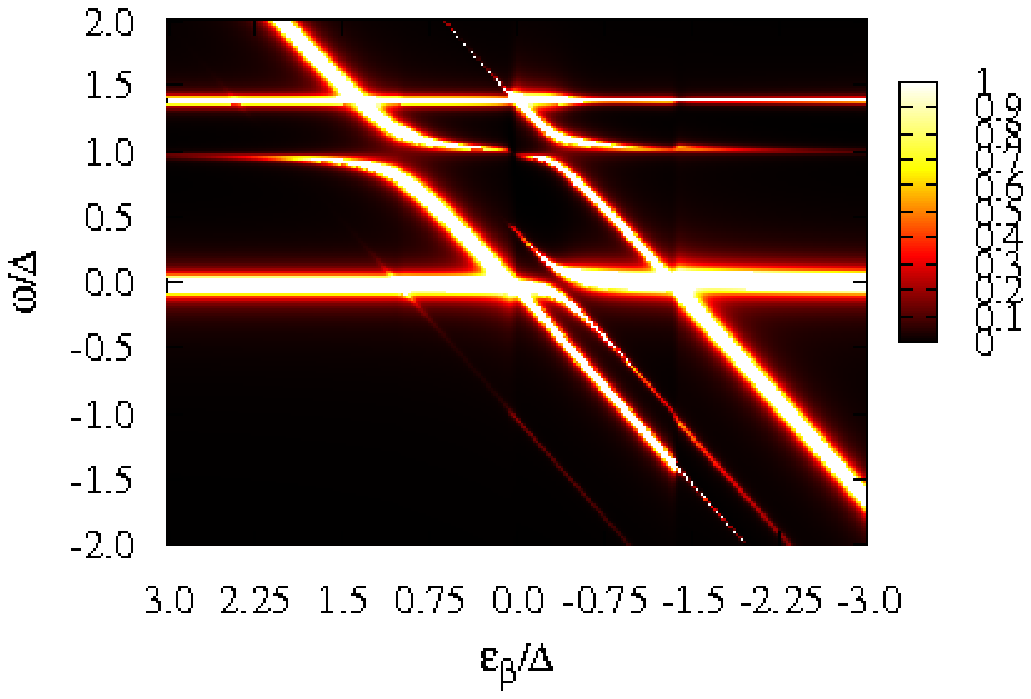}}\\
\hspace*{-6ex} \resizebox{3.0in}{!}{\includegraphics{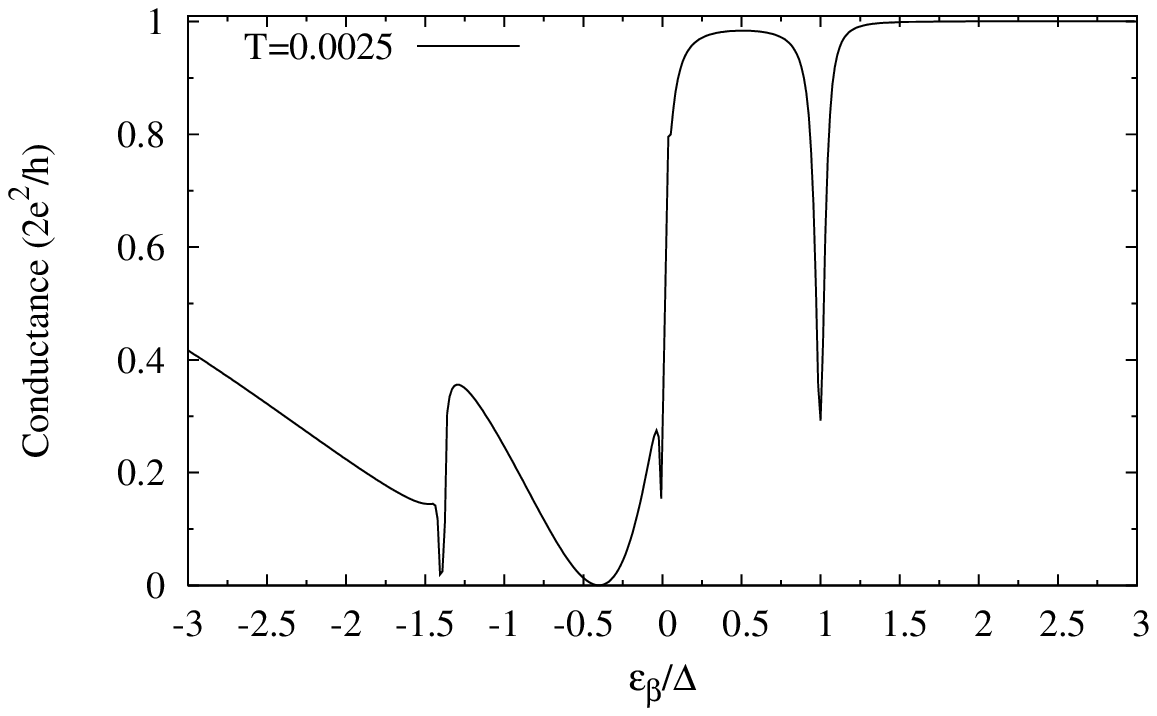}}
\end{tabular}
\end{center}
\caption{\label{cond3} (Color online) Top: Map of the density of states as function of
energy (vertical axis) and  level position $\epsilon_\beta$ (horizontal axis).
Bottom: Conductance as function of energy position $\epsilon_\beta$. Energies in
units of $\Delta = D/8$.  Other parameters are $\hbar\omega_0=1 \cdot \Delta$, 
and $U=1.4 \Delta$, $\tilde t^\prime =\sqrt{2}/5$, $\lambda = 0.2$, and $T =
0.0025 \Delta$. Level $\alpha$ is kept fixed at  $\epsilon_\alpha= -0.001 \Delta$,
essentially at the Fermi level and always contributing to the conductance and
interference. Notice $\epsilon_\beta$ scale runs opposite each other in both panels.}  
\end{figure}

\begin{figure}
\begin{center}
\begin{tabular}{cc}
  \resizebox{3.4in}{!}{\includegraphics{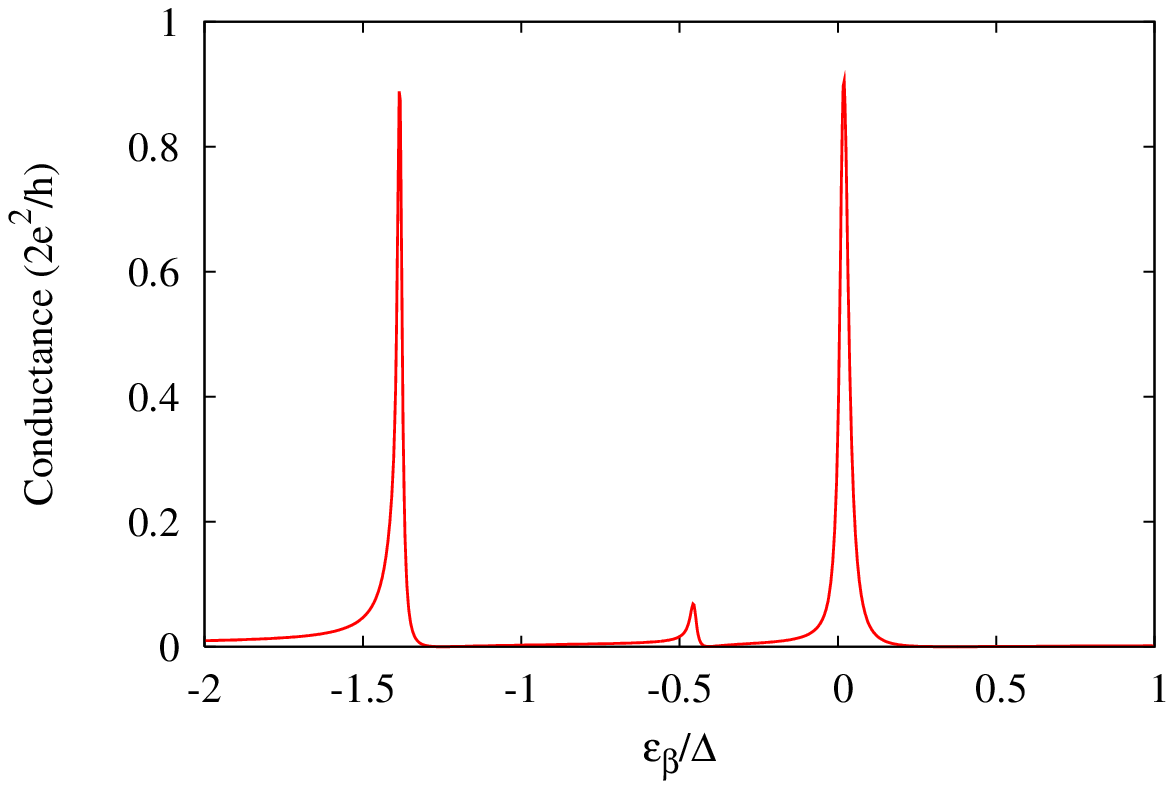}}
\end{tabular}
\end{center}
\caption{\label{cond4} Conductance as function of position of level
$\epsilon_\beta$ for the same case as in Fig.\ \ref{cond3}, but with the 
$\epsilon_\alpha$ level well below the Fermi energy, $\epsilon_\alpha =
-0.1/Delta$, so that it does not contribute to the conductance.
}
\end{figure}

In order to fully probe all the resonance conditions, we fix $\hbar\omega_0$ and
$U$ and  control the energy spacing $\Delta\epsilon$ by gating {\it only one} of the
localized levels. Figure \ref{cond3} shows a color map of the DOS as function
of frequency and energy of the variable level $\epsilon_\beta$. In this plot $\epsilon_\beta$ is shifted down from above to below $\epsilon_F$ as $V_g$ increases; $\epsilon_\alpha$ is kept fixed at a point just below the Fermi energy, $\epsilon_\alpha = -0.001$, so that $\Delta_\epsilon$ also changes with $V_g$.  Since the energy spacing is not constant, we choose instead $\Delta=D/8$ as the energy unit, where $D$ is the bandwidth in the leads. Let us analyze in detail the various peaks in this figure. Following a vertical line, such as $\epsilon_\beta=2.25\Delta$, one finds DOS peaks at $\epsilon_\alpha$, $\epsilon_\alpha+\hbar\omega_0$ and $\epsilon_\alpha+U$ ($\approx 0$, 1, and 1.4), respectively. There is one more peak at $\epsilon_\alpha+U+\hbar\omega_0$ ($\approx 2.4$), which is out of range of the vertical axis in this panel. Note that peaks at both $\epsilon_\alpha \approx 0$ and $\epsilon_\alpha+U$ appear since the site $\alpha$ is in the mixed valence regime. Now following a
horizontal line, at $\omega=-\Delta$, we find peaks corresponding to
the levels $\epsilon_\beta-\hbar\omega_0$, $\epsilon_\beta$,
$\epsilon_\beta+U-\hbar\omega_0$ and $\epsilon_\beta+U$, respectively. One
notices anticrossings of levels whenever one of the levels is a phonon replica, and
crossings when both levels are ``purely'' electronic. The anticrossings reflect a
direct interaction between the electronic levels mediated by phonons. The
crossings appear because the levels interact only indirectly through the leads, an
interaction that is energy dependent.

The bottom of Fig.\ \ref{cond3} shows the conductance of the system. Note the  appearance of antiresonances in the conductance whenever a crossing or anticrossing coincides with the Fermi level. The antiresonances localized near $1$ and $-0.4$ are due to the anticrossings in the density of states discussed above. They result from the absence of states
at the Fermi level. On the other hand the dips near $-1.4$ and $0$ are due
to level crossings. Notice that there is a non-zero contribution to the conductance coming from level $\epsilon_\alpha$, as it is always kept at the Fermi level. The full uninhibited $\alpha$-level conductance is evident for $\epsilon_\beta > 1.5$, where the $\beta$-level is empty.  Notice yet, that the total conductance exhibits dips and even complete vanishing due to the complex pattern of interference arising from competing channels of electrons traversing the molecule through both arms of the structure whenever $\epsilon_\beta$ is near the Fermi level.  In particular, the vanishing conductance at $\epsilon_\beta \simeq -0.5$ appears from the strong level anticrossing lying right at the Fermi level (see upper panel in Fig.\ \ref{cond3}).  

In Fig.\ \ref{cond4} we plot the conductance of the system as function of
$\epsilon_\beta$, where the level $\alpha$ is shifted to a position
below the Fermi energy  ($\epsilon_\alpha = -0.1\Delta$), in such a way
that it essentially does {\em not} contribute to the conductance. As we apply a gate voltage
to change $\epsilon_\beta$, we observe a conductance curve consisting of three
peaks. Two of them,
located near $\epsilon_\beta=0$ and $\epsilon_\beta=-1.4\Delta$ are the usual Coulomb blockade peaks, which are essentially not affected by the EPI.\@ The effect of phonons, however, can be seen most clearly in the appearance of an additional peak near $-0.5\Delta$, due to phonon-assisted processes. This peak arises from the anticrossing of the levels $\epsilon_\alpha$ and $\epsilon_\beta+U-\hbar\omega_0$, opening a new channel for electron transport.

\section{Conclusion}
In summary, we have studied in detail a  diatomic model when both
EEI and EPI are taken into account. We have presented an accurate description of 
the Rabi-assisted tunneling phenomena studying  the Green's functions of the system  in the Coulomb blockade regime above the Kondo temperature. We have obtained well defined resonance conditions for the system. A detailed analysis of
the various resonance conditions and interacting regimes shows that the
most dramatic changes occur for weak EEI ($U,\hbar\omega_0<\Delta\epsilon$), since the EPI is enhanced when the states involved in the phonon emission and absorption are close in energy. These results emphasize the rich
physics involved in phonon-assisted phenomena, which we believe should be
acessible in experiments.  The fact that the phonon-induced resonance conditions are subtly temperature dependent makes the effect strong and sensitive to be probed in experiments.

\begin{acknowledgements}
We acknowledge support from FAPERJ, CNPq (CIAM project) and CAPES in Brazil, as well as from the NSF-IMC grant 0336431 in the US.  
\end{acknowledgements}

% \bibliography{/home/vernek/Documents/Teses/Doutorado/references}

% \end{document}

\end{document}